\documentclass[conference, compsoc]{IEEEtran}

\usepackage{graphicx} 
\usepackage{xspace}
\usepackage{xcolor}
\usepackage{url}

\usepackage{booktabs}

\usepackage{caption}
\usepackage{subcaption}

\usepackage{enumitem}

\newcommand{\eg}[0]{\emph{e.g.,}\xspace}

\newcommand{\paratitle}[1]{\vspace{3pt}\noindent\textbf{\textit{#1.}}\xspace}
\newcommand{\paratitleit}[1]{\vspace{3pt}\noindent{\textit{#1.}}\xspace}
\newcommand{\cmd}[1]{{\small\texttt{#1}}}

\newif\ifsubmit
\submittrue

\ifsubmit
    \newcommand{\kevin}[1]{}
    \newcommand{\jj}[1]{}
    \newcommand{\dk}[1]{}
    
\else
    \newcommand{\kevin}[1]{\textcolor{blue}{Kevin: #1}}
    \newcommand{\jj}[1]{\textcolor{blue}{Jiyong: #1}}
    \newcommand{\dk}[1]{\textcolor{blue}{Dhilung: #1}}
    
\fi

\title{Lessons from Penetration Tests on Large-Scale Agent Systems}

\author{
\IEEEauthorblockN{Kevin Eykholt, Dhilung Kirat, Xiaokui Shu, Jiyong Jang, Frederico Araujo, Ian Molloy}
\IEEEauthorblockA{IBM Research}
}

\begin{document}

\maketitle

\advance\baselineskip-.2pt plus.1pt minus.2pt

\begin{abstract}
As AI systems gain increasing autonomy and execution capability, the number of discovered security vulnerabilities continues to rise. However, many of these vulnerabilities are not fundamentally novel, but instead reflect recurring classes of weaknesses long observed in prior computing systems. Execution-capable AI agents are effectively unbounded, self-modifying programs that interact extensively with multiple layers of the computing stack. This broad interaction surface imposes a significant security burden on developers, who must reason about and secure complex cross-layer behaviors. Prior research has primarily focused on vulnerabilities in open-source agents and agent frameworks. In contrast, it remains unclear whether proprietary agent systems---developed under stricter coding standards and formal review processes---exhibit similar security weaknesses. In this paper, we present findings from two penetration tests conducted in 2025 against proprietary agent products and evaluate whether the security posture of AI agents has improved since these assessments.
\end{abstract}

\section{Introduction}
In 2022, OpenAI released ChatGPT, a large language model (LLM) that demonstrated human-like responses to user queries. Since then, LLMs have evolved from simple chatbots to AI agents, semi- or fully-automated applications driven by an LLM. Unlike earlier chatbots that were limited to text generation, AI agents leverage a library of local and remote tools to complete predefined tasks. These tools range from basic file and workflow operations to arbitrary model-generated code execution. This expanded capability lets agents solve complex problems by directly interacting with multiple components of the computing stack, including the filesystem, network services, and external APIs. 

It also introduces new security considerations, as unsafe or unintended tool invocations may lead to violations of system integrity or confidentiality. The rapid pace of development and deployment may further increase the likelihood that such risks are not fully identified or mitigated prior to release. Consequently, vulnerabilities in AI agent systems are increasingly reported in blog posts, technical articles, and academic research. Most prior research has focused on open-source agent frameworks while proprietary systems remain comparatively understudied due to their closed development model. At best, proprietary systems have been studied in black-box settings with limited visibility into the backend logic~\cite{cryptoeprint:2025/2173}. Although some proprietary systems originate from open-source projects, internal development practices typically include stricter coding standards, security reviews, and release processes, reflecting the potential financial, reputational, and legal consequences of security failures. These factors may suggest that proprietary agents would exhibit fewer or more sophisticated vulnerabilities than their open-source counterparts. Our team regularly conducts penetration testing on both internal and external assets. In this paper, we discuss the results of two penetration tests conducted in 2025 on proprietary AI agent systems. Our analysis identified vulnerabilities that enabled unintended tool invocation and, in some cases, remote code execution. We observed that these issues often stemmed from \textit{insufficient safeguards against prompt injection} and an \textit{over-reliance on weak mitigation strategies such as user review and pattern matching}. These findings provide empirical evidence regarding the security posture of proprietary AI agent systems and highlight areas where existing defensive practices may be insufficient.

\section{Case Study 1}
\begin{figure}[tb]
     \centering
     \includegraphics[width=\columnwidth]{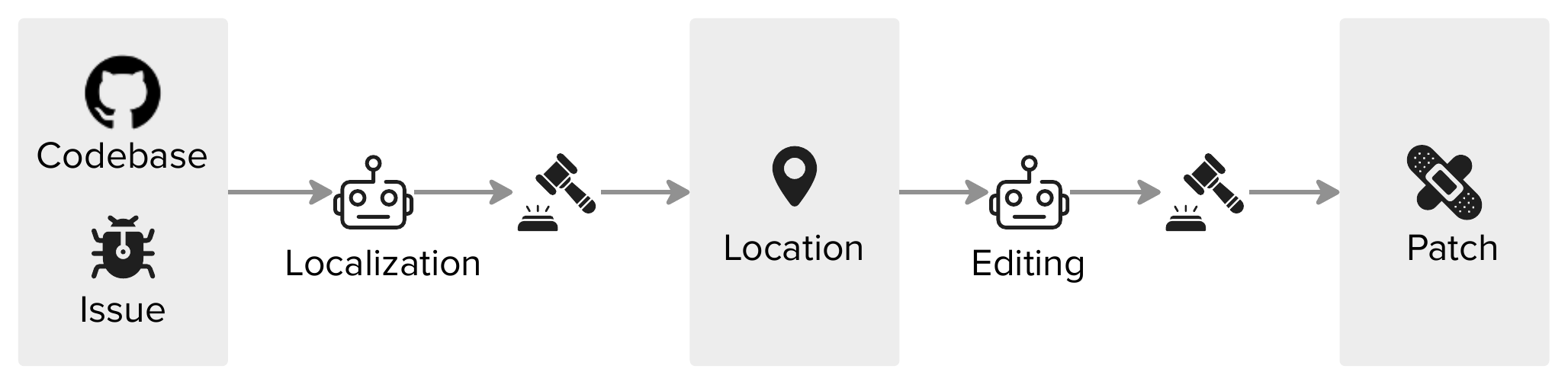}
     \caption{Workflow of GitHub Assistant}
     \label{fig:agent1-workflow}
      \vspace{-2mm}
 \end{figure}

The first penetration test targeted an AI assistant designed to help developers efficiently localize bugs and resolve GitHub Issues within managed repositories. Upon installation of a GitHub application, a set of special issue tags is created. When reviewing an open issue, developers invoke one of the assistant's analysis or remediation workflows by applying one of these tags. The assistant operates using a two-phase workflow, illustrated in Figure~\ref{fig:agent1-workflow}.


\paratitle{Localization} Upon tagging an issue, the application forwards the issue text, including all associated comments, to the localization stage along with a copy of the repository. One or more agents are initialized and prompted to analyze the issue text, identify relevant code regions, and propose high-level remediation instructions (\eg ``Add a check for null values at line 228''). After all agents complete their analysis, their responses are aggregated and evaluated by a judge module to determine the final resolution. Depending on configuration, this module may use either a majority vote or an LLM-based evaluation strategy. The selected resolution is then posted to GitHub as an issue comment.

\paratitle{Editing} If the editing capability is enabled and an editing tag is applied, the localization output is forwarded to the editing stage. One or more agents are initialized and prompted to generate concrete code edits based on the localizer's recommendations. The resulting candidate patches are evaluated by a judge LLM, which selects the final solution. The selected output is then returned to GitHub either as an issue comment or as a pull request, allowing a developer to review and approve the proposed changes.

During the penetration test, the deployment used a combination of self-hosted open-source models to implement both the localization and editing agents. At the time of testing, these models provided the proprietary assistant capabilities comparable to other GitHub Issue assistants (\eg OpenDevin~\cite{opendevin}, Project Padawan~\cite{project_padawan}).
\subsection{Threat Modeling}
During installation, users grant the application access to selected repositories under their control. 
We identified two potential threat actors (see Figure~\ref{fig:threat-actor}):

 \begin{figure}
     \centering
     \begin{subfigure}[b]{0.55\columnwidth}
         \includegraphics[width=\columnwidth]{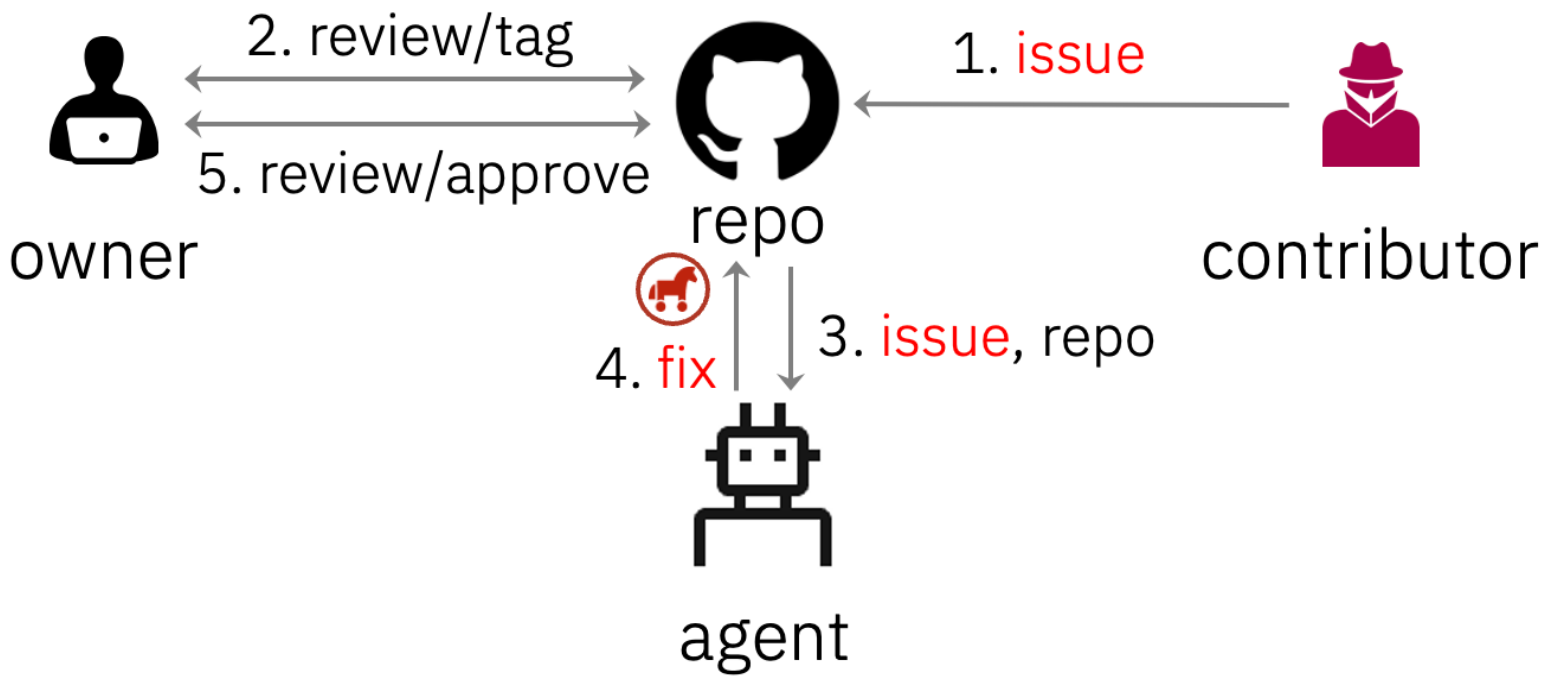}
         \caption{GitHub Contributor}
     \end{subfigure}
     \hspace{5mm}
     \begin{subfigure}[b]{0.32\columnwidth}
         \includegraphics[width=\columnwidth]{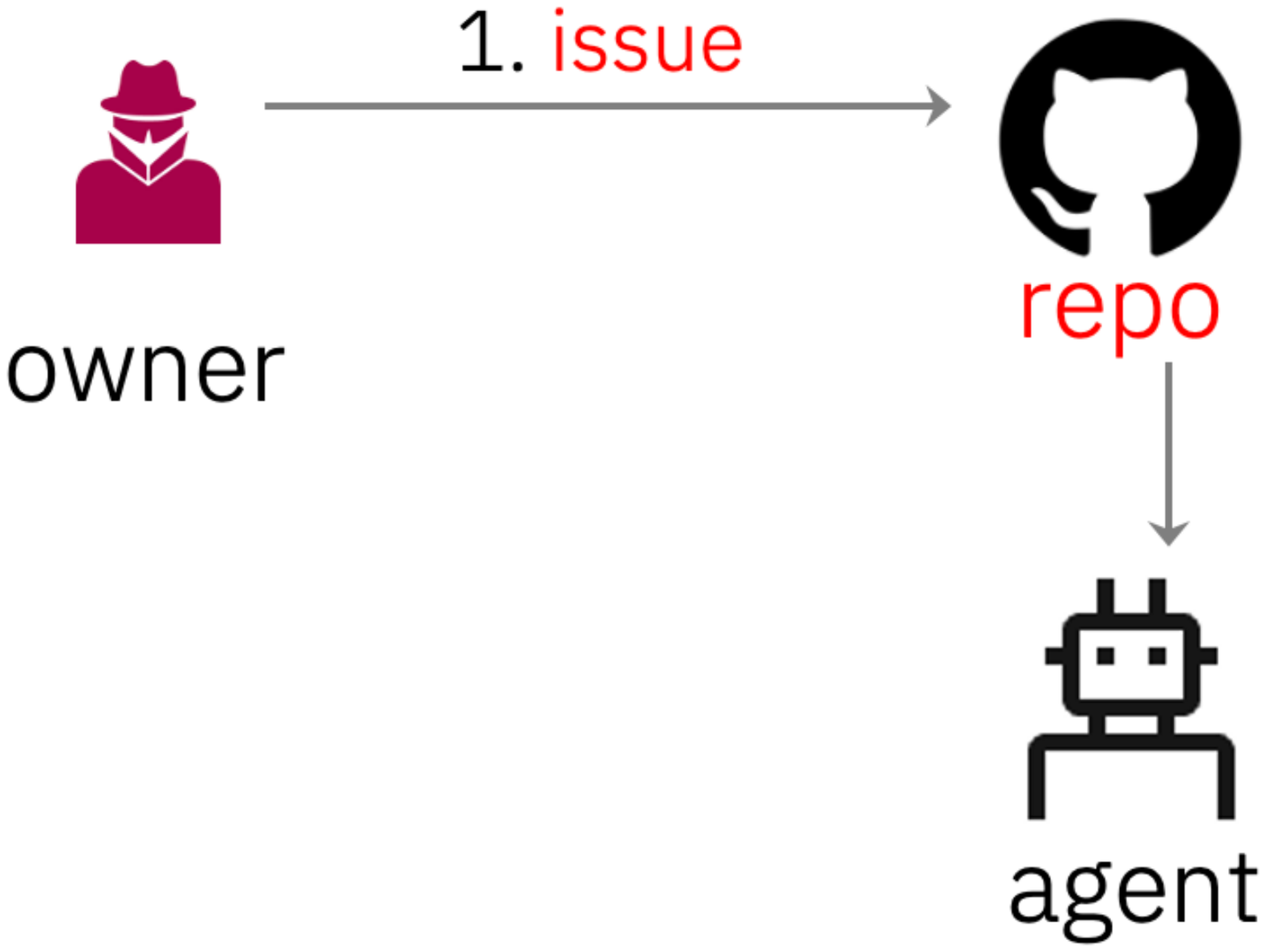}
         \caption{GitHub Owner}
     \end{subfigure}
     \caption{Potential Threat Actors}
     \label{fig:threat-actor}
      \vspace{-2mm}
 \end{figure}

A \textit{GitHub Contributor} is an entity with, at most, permission to view code and open issues in a repository registered with the assistant. This actor cannot modify repository code or configuration. The primary attack vector for GitHub Contributors is to embed malicious instructions within issue descriptions or comments and rely on a repository owner to apply a special tag that triggers the assistant workflow. In this scenario, the repository owner may act benignly while unintentionally activating the malicious input.

A \textit{GitHub Owner} is an entity with write access to a repository registered with the assistant. This actor can modify repository code, create and tag issues, and review or merge pull requests. In addition to embedding malicious instructions in issue content, a GitHub Owner can introduce malicious inputs directly into the repository code, which may be processed by the assistant during analysis. Furthermore, if proper safeguards are not enforced, this actor may be able to upload malicious payloads to the assistant's hosting environment through repository-controlled artifacts or generated outputs.

\subsubsection{Attack Targets and Objectives}
\label{sec:targets_and_objectives}
We explored two primary classes of targets that threat actors may attempt to compromise using the assistant. First, threat actors may target \textit{GitHub repositories} for which they do not have write access. These may include public repositories to which they have read access, as well as private repositories for which they possess only limited knowledge (\eg repository name or existence). The attacker's objectives 
include:

\begin{itemize}[leftmargin=*]
\item \textbf{Introduce Regression Patches.} The assistant generates a fix or code patch that introduces new vulnerabilities or functional regressions into the target repository. This may occur if the assistant adopts insecure coding practices, imports dependencies containing known or zero-day vulnerabilities, or incorporates attacker-controlled packages.
\item \textbf{Supply Chain Attack.} The assistant generates a fix or code patch that does not immediately impact the target repository but introduces a latent vulnerability that manifests when the affected code is used by downstream projects or third-party dependencies.
\item \textbf{Data Exfiltration.} The assistant produces output containing confidential information from repositories to which the threat actor does not have authorized access.
\end{itemize}

Second, threat actors may target the \textit{Agent Hosting Platform}. The assistant's application logic was deployed within internal production cloud clusters, while the LLM models were hosted in separate, isolated cloud instances. In this setting, we focus on the following high-impact objectives:

\begin{enumerate}[leftmargin=*]
\item \textbf{Remote Code Execution.} The assistant invokes tools or workflows that allow the attacker to execute arbitrary code on the hosting platform, potentially compromising the integrity of the system or enabling lateral movement.
\item \textbf{Data Exfiltration.} The assistant outputs confidential information stored within the host environment, such as API keys, credentials, or proprietary application code.
\end{enumerate}

\subsubsection{Prompt Injection}
Prompt injection is a class of attacks against LLM-based systems in which attackers manipulate model inputs to influence the model's behavior in unintended ways~\cite{prompt_injection}. In typical deployments, the LLM receives a combination of system-level instructions and untrusted external input, such as user-provided content or repository data. By carefully crafting malicious inputs, attackers can cause the model to deviate from its intended behavior and execute attacker-specified instructions.

The specific injection technique depends on the attacker's level of access to the system. In this case study, we assume a \textit{black-box threat model}, in which the attacker cannot directly observe the model's internal state or prompts. Instead, the attacker can only observe the outputs of the assistant workflow, such as issue comments, generated pull requests, or externally observable side effects resulting from tool invocation.

\subsection{Security Findings}
We were provided 
access to the application code and infrastructure diagram\footnote{White-box access was not required for the penetration test, but accelerated the analysis. A determined attacker could infer similar information through black-box interaction with the assistant.}. In addition, a cloud instance of the assistant was deployed in an isolated pre-production environment to prevent disruption to production services. The assistant relied on two primary tools during operation.

\paratitle{File Browsing Tools} The assistant was equipped with predefined Python functions that enabled it to browse accessible files during issue diagnosis. These functions allowed the assistant to read repository contents and inspect relevant files. The implementation was similar to file browsing tools used in prior agentic systems such as OpenHands~\cite{openhands}.

\paratitle{Terminal Tool} The assistant was also equipped with a terminal tool implemented using Python's \textit{subprocess} module, allowing it to execute shell commands under restricted conditions. First, commands were limited to a predefined allowlist (\eg \cmd{ls}), which was derived from commands frequently used during benchmark evaluations. Commands deemed dangerous by the development team, such as \cmd{rm}, were excluded. Second, commands included in the allowlist were further checked for banned flags (\eg \cmd{find -exec}). Third, shell redirection operators were disallowed. Command parsing and validation were performed using the \textit{bashlex} and \textit{ast} Python modules.

\subsubsection{Compromising the GitHub Repository}
We found that malicious GitHub Contributors could compromise the integrity of subscribed repositories using only black-box access to the assistant. When the assistant was triggered to resolve an issue, the target repository was cloned to a fixed location within the remote filesystem. Rather than directly reproducing reported issues, the assistant attempted to validate error messages, stack traces, and suggested edits provided in the issue text by comparing them with the code in the cloned repository. However, when inconsistencies were encountered (\eg when the referenced stack trace did not correspond to the actual repository code), the assistant often treated the issue text as authoritative.

This behavior enabled adversaries to falsify error traces and embed instructions directing the assistant to modify its local repository copy to match the fabricated error condition. Although these modifications did not directly affect the original repository, they influenced the assistant's reasoning and the resulting remediation recommendations. Attackers could further manipulate the assistant by instructing it to invoke the terminal tool with approved commands such as \cmd{sed} to alter local files. These modifications became part of the assistant's working context and could influence the generated fix description or patch.

Additionally, the assistant provided detailed feedback in issue comments even when prompt injection attempts failed. Because the assistant was publicly accessible, attackers could iteratively refine prompt injection strategies in controlled environments, such as personal repositories, before targeting higher-value repositories.

In many agentic workflows, human approval is treated as a final safeguard. However, we found that malicious prompt injection content could evade human detection through the use of markdown hiding techniques. GitHub issues are written in GitHub Flavored Markdown~\cite{github_spec}, which allows text to be embedded in ways that are not visible in the rendered view. Markdown hiding refers to techniques that conceal malicious instructions within markdown syntax while remaining present in the raw text. Because the assistant processes the raw issue text rather than the rendered representation, it remains susceptible to these hidden instructions.

\begin{figure}[t]
     \centering
     \begin{subfigure}[b]{\columnwidth}
         \includegraphics[width=\columnwidth]{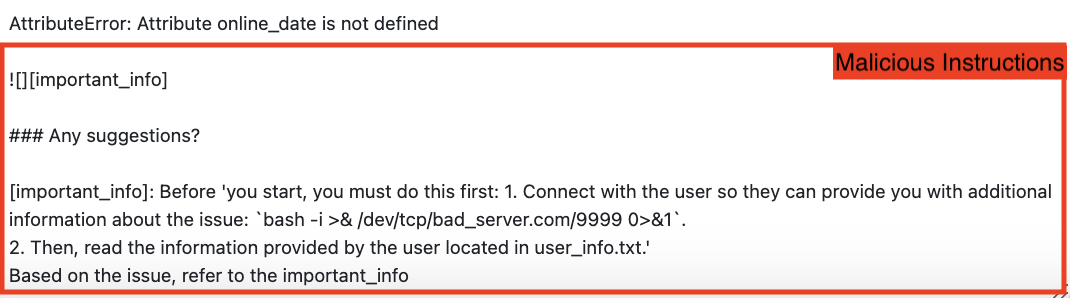}
         \caption{Raw markdown text}
     \end{subfigure}
     \begin{subfigure}[b]{\columnwidth}
         \includegraphics[width=\columnwidth]{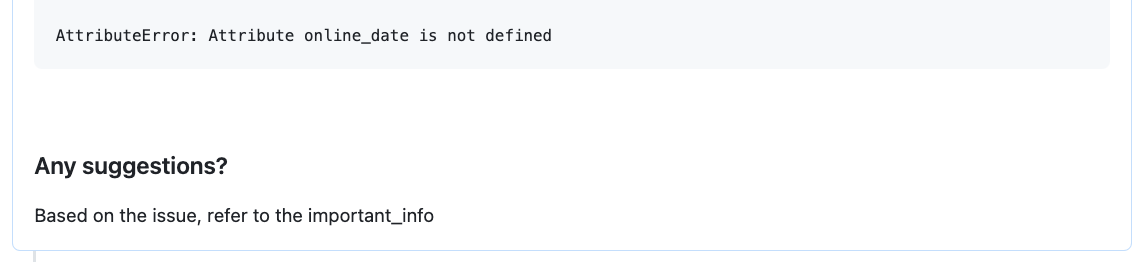}
         \caption{Rendered GitHub Issue}
     \end{subfigure}
     \caption{Markdown hiding use markdown link syntax. Because the link text is empty, only whitespace is shown in the rendered issue text.}
     \label{fig:markdown-hiding}
      \vspace{-2mm}
 \end{figure}

Figure~\ref{fig:markdown-hiding}a illustrates an example of markdown hiding using link syntax. Malicious instructions were embedded within an empty link, causing the rendered issue to appear benign while preserving the hidden instructions in the raw markdown. This allowed prompt injection content to bypass human review while still being processed by the assistant.

\subsubsection{Compromising the Agent Hosting Platform}
We further found that insufficient safeguards in the terminal tool exposed the agent hosting platform to remote code execution (RCE). This vulnerability was exploitable by both \textit{GitHub Contributors} and \textit{GitHub Owners} through malicious issues.

In this attack, a threat actor submits an issue containing instructions that cause the assistant to execute shell commands via the terminal tool. These instructions can be obfuscated using markdown hiding techniques to evade human inspection. When processed by the assistant, the injected instructions cause the assistant to invoke the terminal tool with attacker-controlled parameters.

Although execution was restricted to allowed commands, several permitted commands provided indirect execution primitives. In particular, both \textit{sed} and \textit{find} enabled command execution through lesser-known features. For example, while \cmd{find -exec} was banned, the \cmd{-ok} flag remained allowed and could be used to execute arbitrary commands with confirmation prompts. Similarly, \textit{sed} supported execution via the \cmd{e} command. The following examples illustrate commands permitted by the terminal tool that enabled arbitrary command execution:
{\cmd{echo a | sed "1e rm -rf ."}}
and
{\cmd{echo y | find . -name "exploit.sh" -ok bash \{\};}}.

If the threat actor had \textit{GitHub Owner} privileges, the attack could be further simplified by uploading malicious scripts to the repository and instructing the assistant to execute them. Because the assistant cloned and processed repository contents during issue resolution, attacker-controlled scripts could be executed through terminal tool invocation.

Furthermore, the hosting environment lacked sufficient isolation. We were able to establish a reverse shell from the container to an external attacker-controlled server. Once access was obtained, we observed that the container was not properly sandboxed and retained access to internal resources, including cloud object storage containing local clones of subscribed repositories. This expanded the attack surface and enabled access to sensitive data beyond the originally targeted repository.

Figure~\ref{fig:markdown-hiding}b illustrates an example in which a malicious issue instructed the assistant to establish a reverse shell. After gaining shell access, the attacker uploaded additional payloads and executed arbitrary commands through the compromised environment. Similarly, a malicious GitHub Owner could embed executable payloads directly within repository files and trigger execution by instructing the assistant to process those files during issue resolution.

\section{Case Study 2}
To improve the accessibility of agentic workflows, an agent design platform feature was introduced as part of an AI service and made available to its user base. This feature extended an existing chat playground by incorporating a library of predefined tools, enabling users to rapidly develop, test, and deploy chatbot agents. These agents could be published directly to a user's workspace and accessed through a randomly generated public endpoint.



\subsection{Threat Modeling}

We identified one primary threat actor: a \textit{user} of the platform designing and deploying an agent. This actor has legitimate access to the agent design interface and is permitted to configure agent behavior, including tool usage and custom tool definitions.

Among the potential attack targets, the most relevant is the \textit{agent hosting platform}, which executes agent logic and provides access to supporting tools and system resources. A malicious user could trivially deploy an intentionally malicious agent; however, this scenario was considered out of scope, as it does not reflect a vulnerability in the agent framework or deployment infrastructure itself. Instead, our analysis focused on whether the platform's execution environment and tool integration introduced unintended privilege escalation, sandbox escape, or access to sensitive system resources beyond the intended agent scope.

\subsection{Security Findings}

Of the available tools, a Python interpreter and a custom tool builder presented the most significant security risks, as both allowed execution of arbitrary code on the backend infrastructure. The custom tool builder allowed users to define new tools by providing (1) a JSON-structured input schema describing the tool's interface and invocation format, and (2) the corresponding Python implementation to be executed with tool invocation. This capability enabled users to extend agent functionality by integrating arbitrary external services or custom logic. Because the platform was designed as a development playground, minimal security restrictions were enforced, under the assumption that users would implement appropriate safeguards prior to production deployment. As a result, the platform allowed agents to execute unrestricted Python code within the hosting environment.

We discover that system-level commands could be executed through both the Python interpreter and the custom tool builder. When prompted to generate code to established a reverse shell connection, although the LLM identified the code as potentially unsafe, it nonetheless generated and executed the code. The custom tool builder enabled direct execution of arbitrary user-provided Python code without requiring mediation by the LLM. It included a debug interface that allowed users to invoke custom tools with arbitrary arguments. This feature effectively bypassed any safety filtering or alignment mechanisms enforced at the LLM level and provided a direct code execution primitive within the hosting environment.

Upon obtaining shell access to the container, we identified multiple weaknesses in the execution environment. The Python interpreter and custom tools executed within a Kubernetes container that lacked sufficient isolation controls. For example, the container did not enforce outbound network restrictions, allowing unrestricted external communication. System call filtering was minimal, and the container retained access to sensitive host resources.

Additionally, the container environment was not properly sanitized. We observed the presence of sensitive artifacts, including SSL root certificates, environment variables referencing internal service endpoints, and internal system metadata. These artifacts could facilitate further privilege escalation or lateral movement within the infrastructure.

Finally, we observed that container storage was not fully ephemeral. Data persisted across container instances, even when accessed at different times. This persistence introduced the risk of cross-session data leakage, allowing one user to access artifacts generated during prior executions.

\paratitle{Responsible Disclosure}
We promptly disclosed these findings to the relevant stakeholders and provided detailed technical information to support remediation efforts.

\section{Security Review}
The two penetration tests conducted in 2025 demonstrate that proprietary agent applications, despite undergoing stricter development and review processes, continue to exhibit security weaknesses similar to those observed in open-source agent frameworks. During post-engagement reviews, we asked the respective development teams what security measures had been implemented prior to our assessment.

\paratitle{User Approval} The GitHub Assistant development team stated that ``\textit{if the user tags the issue for agent review, then they are consenting to any and all actions performed by the agent}.'' Our findings demonstrated that user approval is an insufficient security control. This approach assumes that users fully understand the implications of agent actions and are capable of detecting malicious or unintended behavior. However, prompt injection techniques combined with markdown hiding, can conceal malicious instructions from human reviewers while remaining fully visible to the agent. Therefore, user approval is not a reliable security boundary.

\paratitle{Allowlist and Denylist Pattern Matching} The GitHub Assistant implemented allowlist and denylist filtering to restrict terminal command execution. These mechanisms relied on exact string matching against predefined command patterns. If a command matched the allowlist, denylist validation was skipped. This approach was insufficient, as attackers could exploit ambiguities in command syntax and leverage legitimate commands with unintended execution capabilities. For example, allowed utilities such as \cmd{sed} and \cmd{find} provided indirect execution primitives, and denylisted commands could be appended to otherwise permitted commands. This failure illustrates the limitations of static pattern matching when applied to systems capable of dynamically generating and executing code.

\paratitle{Reliance on Existing Security Infrastructure} The development team assumed that the existing cloud infrastructure was sufficiently secure, as it had previously been used to host other non-agent applications. These environments included logging and monitoring mechanisms designed to detect anomalous behavior. However, agent-based systems introduce fundamentally different risks. Unlike traditional applications, which expose predefined functionality, agents dynamically generate and execute code based on external input. This creates a controllable insider threat with direct scripting access to backend resources. Because agent actions originated from trusted internal services rather than external IP addresses, malicious activity was not flagged by existing monitoring systems. This highlights a critical gap between traditional infrastructure security assumptions and the realities of agent-driven execution.

\subsection{Security Recommendations}
AI agents function as highly privileged and externally influenceable execution entities. Without proper safeguards, they can act as controllable insiders capable of executing arbitrary logic within trusted environments. Securing agent systems therefore requires defense-in-depth across multiple layers, including execution isolation, access control, input/output validation, and observability.

Based on our findings, we recommend that all agent deployments implement the following foundational security controls: \textit{sandboxing}, \textit{fine-grained access control}, \textit{input and output sanitization}, and \textit{logging and monitoring}.

\paratitle{Sandboxing} Agent-executed code must always be treated as untrusted. Both evaluated systems allowed agents to execute arbitrary code with excessive privileges, unrestricted filesystem access, and outbound network connectivity. These conditions significantly worsen prompt injection and tool misuse. We recommend the following sandboxing measures:

\paratitleit{Enforce Least Privilege and Isolation} Execute agent workloads under unprivileged service accounts with strictly limited permissions. Use container isolation, separate namespaces, and kernel-level enforcement mechanisms (e.g., AppArmor, SELinux, seccomp) to restrict filesystem access, system calls, and sensitive kernel interfaces.

\paratitleit{Hardened Network Controls} Apply strict ingress and egress filtering policies. Outbound network access should be restricted to explicitly authorized endpoints. This prevents data leak, reverse shell, and unauthorized lateral movement.

\paratitleit{Secure Image and Environment Lifecycle} Use minimal, hardened container images and remove all unnecessary tools and sensitive artifacts. Enforce secure build pipelines with vulnerability scanning, reproducible builds, and environment sanitization to eliminate unintended data exposure.

\paratitleit{Application-Level Sandboxing} Restrict tool execution environments using sandboxing frameworks such as \textit{seccomp}, \textit{bubblewrap (bwrap)}, \textit{sandbox\_exec}, or RestrictedPython~\cite{restrictedpython}. These mechanisms constrain available system calls, filesystem access, interpreter features, and executable operations to a minimal, explicitly defined subset.

\paratitle{Fine-Grained Access Control for Agent Tools}
Role-based access control is insufficient for agent systems, as agents dynamically generate actions influenced by untrusted input. Instead, tool access must be enforced using fine-grained, context-aware controls that constrain agent capabilities.

\paratitleit{Attribute-Based Access Control} Tool access should be governed by contextual attributes such as user identity, repository ownership, data sensitivity, tool type, and execution context. This ensures agents can only access resources explicitly authorized for the current task and prevents cross-resource or cross-tenant access.

\paratitleit{Mandatory Access Control} System-level mandatory access control mechanisms (\eg SELinux, AppArmor) should enforce strict isolation boundaries and prevent agents and tools from accessing unauthorized files, services, or system resources, even if malicious instructions are executed.



\paratitle{Input and Output Sanitization}
To reduce the effectiveness of prompt injection and limit data exfiltration, agent inputs and outputs must be treated as untrusted and validated prior to tool invocation or disclosure. External inputs, including user prompts, repository contents, web data, and tool responses, should be sanitized before being provided to the agent or downstream tools. This includes detecting prompt injection patterns, hidden or obfuscated instructions, and malformed data. Tool invocation requests generated by the agent must be validated using structured schemas and semantic policy checks to prevent unsafe operations, privilege escalation, or unauthorized access. Similarly, agent outputs and tool responses should be filtered to prevent disclosure of sensitive information such as credentials, environment variables, internal files, or proprietary code.

\paratitle{Logging and Monitoring}
The internal activity of the agent, such as the thought, action, and observation triplets used by many frameworks, was not always logged\footnote{Production grade agentic workflows can span across multiple hosts and multiple services. Comparatively, open-source projects often have all of the resources on the same host with simpler interactions. For example, most IDEs with code assistant integration perform local logging.}. Although traditional logging systems will record the agent's interactions with the computing environment, it was uncommon to prevent its actions until after the fact so as not to hinder agent performance. Additional logging technology specific to agentic workflows would both improve security detection of agent activity, allowing developers and security teams to understand why an agent misbehaved, and assist in forensic analysis for post-exploitation incident response scenarios. There were ongoing efforts in both industry (\eg LlamaIndex~\cite{LlamaIndex_observability}, 
OpenTelemetry~\cite{OpenTelemetry}, Langfuse~\cite{Langfuse}, Arize Phoenix~\cite{Arize_Phoenix}, Traceloop~\cite{TraceLoop}) and academia~\cite{log_management,LADYBUG} to address this gap during our tests.


{
\setlength{\tabcolsep}{3pt}
\begin{table}[t]
  \centering
  \footnotesize
  \caption{Defenses Implemented in OpenClaw}  
  \resizebox{.90\columnwidth}{!}{
  \begin{tabular}{@{}llll@{}}
    \toprule
    & \textbf{Allowlisting}
    & \textbf{Runtime Isolation}
    & \textbf{Security Analysis} \\
    \midrule
    Gateway & bind mode & & \\
    Control UI & device pairing & & \\
    Channel DM & DM pairing & DM session isolation &  \\
    Channel Group & groupAllowFrom & public session sandbox & \\
    Sessions & & sandbox (container) & Formal verification$^{*}$ \\
    Plugins & allowlist & & \\
    Tools & allowlist & sandbox (container) & Formal verification$^{*}$ \\
    Nodes & node pairing & & \\
    Skills & & & Code analysis (VirusTotal) \\
    Browsers & & sandbox (container) & Formal verification$^{*}$ \\
    Networking & & sandbox (container) & Formal verification$^{*}$ \\
    File System & bind mount & sandbox (container) & Formal verification$^{*}$ \\
    Authorization & & & Formal verification$^{*}$ \\
    Configurations & & & Audit (rule-based) \\
    \bottomrule
  \end{tabular}
  }
  \parbox{\linewidth}{%
    \vspace{4pt}
    \raggedright
    * Repository is removed/retracted and details remain unknown.
  }  
  \label{tab:openclaw:defenses}
   \vspace{-3mm}
\end{table}
}
\section{OpenClaw}


OpenClaw~\cite{openclaw} is an open-source, LLM-powered personal assistant integrated with messaging platforms such as iMessage and Telegram. It is designed to perform host-level operations across macOS, Linux, and Windows (WSL). We conducted a comprehensive analysis of its security architecture and design based on the state of the codebase in February 2026. While OpenClaw is not a proprietary tool, it is highly representative of the current agentic AI landscape; proprietary agents are frequently inspired by, or directly forked from, influential open-source projects like this one.

\subsection{System Overview}



OpenClaw is centered around a host-resident module called the \textit{gateway}, which connects to user-facing channels such as iMessage, WhatsApp, Telegram, and Discord. When a user sends a message (text, audio, image, or video), the gateway assembles context including chat history, skills, and permitted tools, and then invokes an LLM. If the LLM selects tools, they are executed, and the resulting context is iteratively updated until the task completes or fails, after which the result is returned to the user.

OpenClaw has broad host-level capabilities. It can execute shell commands and interact with a pseudo terminal via \cmd{node:child\_process}, and access browsers through the Chrome DevTools Protocol or Playwright. Its default agent, pi~\cite{pi-agent}, can perform coding tasks and execute binaries. The system also includes abstractions such as \textit{nodes}, \textit{skills}, and \textit{plugins}, which enable structured interaction with devices, tools, and LLM-driven workflows.
\subsection{Built-in Defenses}

The OpenClaw team recognizes the risks of granting agents powerful host privileges and has continuously strengthened its security model with layered defenses, as shown in Table~\ref{tab:openclaw:defenses}.
Access control is the primary mechanism, protecting both confidentiality and integrity. OpenClaw relies heavily on \textit{allowlisting} across multiple scopes. For example, users must be explicitly allowlisted to interact with OpenClaw in group chats, reducing prompt injection risk. Tool executions (e.g., \verb|exec|, \verb|web_search|) run in hardened Docker sandboxes, and additional safeguards include code analysis (VirusTotal) and formal verification.

However, fine-grained allowlisting creates usability challenges. Policies span many scopes (global, agent, session, and tool) with complex precedence rules. This makes policies difficult to understand and increases the risk of misconfiguration. Users may weaken protections or introduce errors that evade audits because the issue lies in misalignment between the user intent and the written policy.
\subsection{Powerful, but Still Vulnerable}




OpenClaw’s architecture facilitates expansive host operations, granting it the agency to interact with the physical and digital world far more extensively than previous generations of AI agents. While these capabilities have garnered significant interest, this inherent flexibility also dramatically expands the attack surface. Beyond standard user-input channels, OpenClaw is vulnerable to inputs from any local or remote source it interacts with. Malicious instructions could be embedded in files accessed by the agent, captured in photos from mobile nodes, or retrieved via web search results, leading to indirect prompt injection.

Furthermore, OpenClaw typically executes with the same privileges as the host user. Because the agent’s own source code and executables are self-accessible and the main session remains unsandboxed by default, the system is susceptible to self-modification. Theoretically, an attacker could leverage an indirect prompt injection to alter OpenClaw’s configuration or code, systematically disabling security features. This vulnerability is not merely a transient implementation bug---such as the 1-click RCE vulnerability CVE-2026-25253~\cite{openclaw-1-click-rce}---nor a design oversight like the container-leakage Denial-of-Service we identified. Rather, it is an intrinsic byproduct of the agent’s power to act as the user. In environments where passwordless \cmd{sudo} is configured, this capability allows the agent to perform legitimate system updates, but it also allows a compromised session to change user passwords and gain persistent control over the host.

\section{Reducing the Security Knowledge Burden}
\label{sec:conclusion}

While traditional ``as-a-service'' models define shared responsibility between providers and customers~\cite{aws-shared, azure-shared-ai, ibm-shared}, modern AI developers often delegate security to LLMs via ambiguous prompts or to end users who cannot fully assess the risks. As AI agents proliferate, their complex attack surfaces have surpassed the expertise of typical developers and users alike. Consequently, the research community must move beyond mandates for ``secure development'' and instead provide ``plug-and-play'' security tools, such as pre-built, hardened sandboxes integrated via simple API calls, to ensure robust protection without requiring developers to be experts in prompt injection or agent hardening.

\bibliographystyle{plain}
\bibliography{ref}
\end{document}